\documentclass[twocolumn,aps,prb,superscriptaddress]{revtex4-1}
\usepackage{graphics,bm,graphicx}
\usepackage{amssymb}
\usepackage{epsfig}
\usepackage{epsf}
\usepackage{xcolor}

\def\Im{{\rm Im}}
\def\be{\begin{equation}} \def\ee{\end{equation}}
\def\beq{\begin{eqnarray}} \def\eeq{\end{eqnarray}}

\def\nn{\nonumber}

\begin{document}

\title{Cooperative Effects of Strain and Electron Correlation in Epitaxial VO$_2$ and NbO$_2$}

\author{Wei-Cheng Lee}
\email{wlee@binghamton.edu}
\affiliation{Department of Physics, Applied Physics, and Astronomy, Binghamton University, Binghamton, New York, 13902, USA}

\author{Matthew J. Wahila}
\affiliation{Department of Physics, Applied Physics, and Astronomy, Binghamton University, Binghamton, New York, 13902, USA}

\author{Shantanu Mukherjee}
\affiliation{ Department of Physics, Indian Institute of Technology Madras, Chennai-600036, India}

\author{Christopher N. Singh}
\affiliation{Department of Physics, Applied Physics, and Astronomy, Binghamton University, Binghamton,  New York, 13902, USA}

\author{Tyler Eustance}
\affiliation{Department of Physics, Applied Physics, and Astronomy, Binghamton University, Binghamton,  New York, 13902, USA}

\author{Anna Regoutz}
\affiliation{Department of Materials, Imperial College London, London SW7 2AZ, UK} 

\author{H. Paik}
\affiliation{Department of Materials Science and Engineering, Cornell University, Ithaca, New York 14853-1501, USA}

\author{Jos E. Boschker} 
\affiliation{Leibniz-Institut f\"ur Kristallz\"uchtung, Max-Born-Stra{\ss}e 2, D-12489 Berlin, Germany}

\author{Fanny Rodolakis}
\affiliation{Argonne National Laboratory, 9700 South Cass Avenue, Argonne, Illinois, 60439, USA}

\author{Tien-Lin Lee}
\affiliation{Diamond Light Source, Harwell Science and Innovation Campus, Didcot OX11 0DE, UK} 

\author{D. G. Schlom}
\affiliation{Department of Materials Science and Engineering, Cornell University, Ithaca, New York 14853-1501, USA}

\author{Louis F. J. Piper}
\affiliation{Department of Physics, Applied Physics, and Astronomy, Binghamton University, Binghamton, New York, 13902, USA}

\date{\today}

\begin{abstract}
We investigate the electronic structure of epitaxial VO$_2$ films in the rutile phase using density functional theory combined with the slave-spin method (DFT+SS). In DFT+SS, multi-orbital Hubbard interactions are added to a DFT-fit tight-binding model, and slave spins are used to treat electron correlations. We find that while stretching the system along the rutile $c$-axis results in a band structure favoring anisotropic orbital fillings, electron correlations favor equal filling of the $t_{2g}$ orbitals. These two distinct effects cooperatively induce an orbital-dependent redistribution of the electron occupations and spectral weights, driving strained VO$_2$ toward an orbital selective Mott transition (OSMT).  The simulated single-particle spectral functions are directly compared to V L-edge resonant X-ray photoemission spectroscopy of epitaxial 10 nm VO$_2$/TiO$_2$ (001) and (100) strain orientations. Excellent agreement is observed between the simulations and experimental data regarding the strain-induced evolution of the lower Hubbard band.  Simulations of rutile NbO$_2$ under similar strain conditions are performed, and we predict that an OSMT will not occur in rutile NbO$_2$. Our prediction is supported by the high-temperature hard x-ray photoelectron spectroscopy (HAXPES) measurement on relaxed NbO$_2$ (110) thin films with no trace of the lower Hubbard band. Our results indicate that electron correlations in VO$_2$ are important, and can be modulated even in the rutile phase before the Peierls instability sets in.
\end{abstract}

\maketitle
\section{Introduction}
Enormous efforts have been made to understand vanadium dioxide (VO$_2$), which exhibits a metal-to-insulator transition (MIT) near room temperature.\cite{morin1959} The MIT in VO$_2$ is accompanied by a structural transition from the rutile R (metallic) to the monoclinic M$_1$ (insulating) crystal structure, and it is understood that both Mott (electron correlation)\cite{mott1975,imada1998} and Peierls (structural distortion)\cite{goodenough1971} physics play essential roles in this transition. A large amount of theoretical research has been performed to investigate the electronic structures of VO$_2$ using the local density approximation (LDA), LDA + U, LDA + DMFT, quantum Monte Carlo, etc.. However, the role of electron correlations in driving the MIT is still under debate, due to the fact that the Mott insulating phase cannot be disentangled from the Peierls instability. \cite{eyert2002,haverkort2005,koethe2006,biermann2005, weber2012,zheng2015,brito2016} 

Recent progress in strain-engineering VO$_2$ grown by molecular beam epitaxy on TiO$_2$ substrates has opened the possibility of modulating electron correlation with strain before the Peierls instability sets in. \cite{paik2015, quackenbush2015,mukherjee2016,quackenbush2016,laverock2012} Since TiO$_2$ remains in the rutile crystal structure at all temperatures, and has a $c$-axis lattice constant approximately $3.8\%$ longer than bulk VO$_2$, the strain can be engineered by choosing the thin film growth direction. If the growth direction is along the rutile $c$-axis, denoted as VO$_2$ (001), the strain is small and the corresponding VO$_2$ thin film behaves like bulk VO$_2$.\cite{quackenbush2013} On the other hand, if the growth direction is perpendicular to the rutile $c$-axis, e.g., (100) and (110), a large strain is induced due to significant elongation of the $c$-axis lattice constant. Recent evidence of a strain-induced orbital selective Mott state was reported from hard x-ray photoelectron spectroscopy (HAXPES) and  x-ray absorption spectroscopy (XAS) experiments in epitaxial (100) and (110) films by Mukherjee and Quackenbush et al.\cite{mukherjee2016,quackenbush2016} The reports revealed how electron correlation can be strain-enhanced in VO$_2$ thin films, but a full theoretical description of the interplay between strain and electron correlation remained incomplete.

In this paper, we employ density functional theory implemented with the slave spin method (DFT+SS) to investigate the electronic structure of VO$_2$ thin films under strain. We focus on the rutile phase and study the phase diagram as a function of doping, strain, and correlation strength. We find that while the band structure is modified by the elongation of the $c$-axis driving more electrons into the $d_{\parallel}$ band, electron correlation has the opposite effect, and tends to distribute electrons more evenly among the three $t_{2g}$ bands. The interplay between these two effects results in orbital-dependent redistributions of the electron occupations and spectral weights, which pushes the strained VO$_2$ toward an orbital selective Mott transition (OSMT). Our theoretical models are in excellent agreement with V L-edge resonant x-ray photoemission spectroscopy of both (001) and (100) epitaxial 10 nm VO$_2$/TiO$_2$, which highlights the evolution of the lower Hubbard band as a function of strain. We have further employed the same theoretical approach to study rutile NbO$_2$ under similar strain conditions, and conclude that no OSMT can be found for comparable strain regimes. 
Our conclusion is supported by the high-temperature hard x-ray photoelectron spectroscopy (HAXPES) measurement on relaxed NbO$_2$ (110) thin film with no trace of the lower Hubbard band. Our results demonstrate that electron correlation has significant effects on the physical properties of VO$_2$ even in the rutile phase, and can be modified dramatically by strain engineering.

\section{General discussion}
\begin{figure}
\includegraphics[width=3.4in]{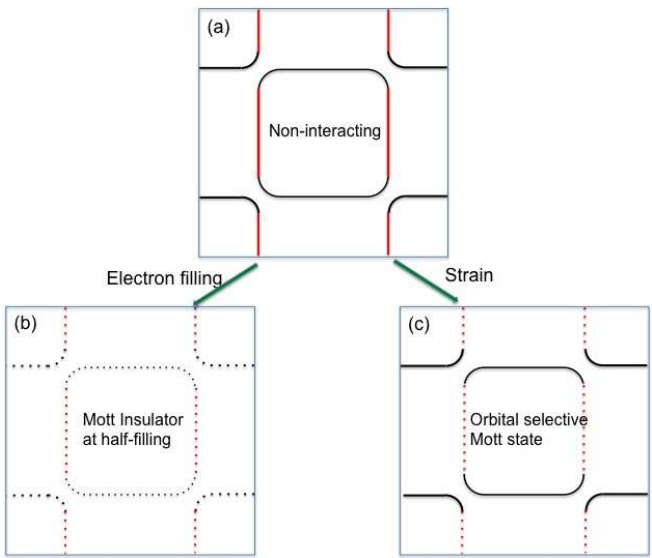}
\caption{\label{fig:OSMT-2o} Schematic illustrations of the Mott insulator and the orbital selective Mott state. (a) The Fermi surface of a non-interacting generic two-orbital system with 
$d_{xz}$ and $d_{yz}$. The red (black) color represents larger $d_{xz}$ ($d_{yz}$) component at these momenta. (b) At half-filling ($n=2$), the Fermi surface completely vanishes in the Mott insulating state. (c) In the orbital selective Mott state, only parts of the Fermi surface associated with one particular orbital vanish, exhibiting an interesting 'Fermi arc' profile. This can be achieved by strain engineering even at non-integer fillings.}
\end{figure}

Mott physics can be generally described by a Hamiltonian containing two terms: $H = H_{t} + H_U$.
$H_{t}$ is the kinetic energy, typically obtainable from band structure calculations, and $H_U$ is the multi-orbital Hubbard interaction that provides a Coulomb penalty 
for two electrons on the same atom. 
In the single band case, if the system is at 'half-filling', which refers to the electron density being 1 electron per site ($n=1$), the transition from metallic to Mott insulating state can be tuned by varying the ratio of $U/W$, where $U$ is the Hubbard on-site Coulomb energy, and $W$ is the bandwidth.

The physics of "Mottness" is even richer in a multi-orbital system. The half-filling condition for the Mott insulating state becomes $n=N_o$, where $N_o$ is the number of orbitals, but there could exist a new state beyond the standard Mott insulator.
Let's consider a generic two-orbital system with $d_{xz}$ and $d_{yz}$ whose typical Fermi surface is plotted in 
Fig. \ref{fig:OSMT-2o}(a). At half-filling ($n=2$), the Mott insulating state can occur with large enough on-site Coulomb interaction, and the Fermi surface will be completely gapped out, as shown in 
Fig. \ref{fig:OSMT-2o}(b). As the electron density moves away from half-filling, the correlation effect becomes weaker, and it is generally expected that Mott physics becomes much less important. 
However, strain engineering facilitates the OSMT mechanism as explained below.

Let's assume that $n_{xz,yz}$ and $W_{xz,yz}$ are the electron density and bandwidth of orbitals $d_{xz}$ and $d_{yz}$. The total electron density is just 
$n=n_{xz} + n_{yz}$,
and the correlation effect on each orbital can be characterized by the ratio of $U/W_{xz,yz}$. The key part of this mechanism is that the orbital wavefunctions are highly anisotropic in real space, thus the bandwidths can be modulated quite differently in different orbitals under the same
tensile strain. In this example of the two-orbital system, the $d_{xz}$ ($d_{yz}$) orbital has a much larger wavefunction overlap along the $x$ ($y$) direction. 
If a tensile strain is applied along the $x$ direction, the lattice constant along $x$ direction will be elongated, which reduces $W_{xz}$ more significantly than $W_{yz}$. This orbital-dependent bandwidth reduction due to tensile strain leads to two important consequences. First, it is entirely possible to put more electrons in the $d_{xz}$ orbital.  Second, the correlation effect is increased only in the $d_{xz}$ orbital, expressed by the ratio of $U/W_{xz}$. As a result, even if the system is not strictly half-filled originally, the application of the strain along $x$ direction could drive the system to a state where only the $d_{xz}$ orbital is Mott insulating ($n_{xz}\sim 1$ with larger $U/W_{xz}$) while the $d_{yz}$ orbital remains weakly-correlated ($n_{yz}$ away from $1$ with smaller $U/W_{yz}$). In this new state of matter, namely the orbital selective Mott state, the Fermi surfaces will be disconnected, exhibiting an interesting 'Fermi arc' profile shown in Fig. \ref{fig:OSMT-2o}(c).

The mechanism described above is quite general for a system with multiple orbitals near the Fermi surface together with strong Hubbard interactions.
\cite{anisimov2002,liebsch2003,liebsch2004,fang2004,koga2004, demedici2005}
In the following sections, we present our theoretical study and supporting experimental data for the case of VO$_2$ thin films. 
We also perform the same calculations for rutile NbO$_2$ under similar strain conditions. 
Our results show that the OSMT occurs in VO$_2$ thin films, but not in NbO$_2$, which can be attributed to the fact that NbO$_2$ has much larger bandwidths in general compared to VO$_2$.

\section{Theoretical Methodology}
We solve the Hamiltonian expressed as
\be
H = H_{TB} + H_U.
\ee
$H_{TB}$ is the tight-binding model composed of the $d$-orbtials of the $V$ atom as well as the $p$ orbitals of the $O$ atom, and the hopping parameters are determined by
fitting the band structure calculated by density functional theory using Wannier90\cite{wannier90}.
The DFT calculations are performed with full potential linear augmented plane waves plus local orbitals (FP-LAPW+lo) and the Perdew-Burke-Ernzerhof generalized gradient approximation (PBE-GGA) provided in the WIEN2k code.\cite{wien2k}
The electron correlations are treated by the multi-orbital Hubbard term $H_U$ on the $d$ orbitals of the $V$ atom defined as
\beq
H_U &=& U\sum_{i\alpha} \hat{n}_{i\alpha\uparrow}\hat{n}_{i\alpha\downarrow}+ \frac{U'}{2}\sum_{i,\alpha\neq\beta} \sum_\sigma \hat{n}_{i\alpha\sigma}\hat{n}_{i\beta,\bar{\sigma}}\nn\\
&+&\frac{U'-J}{2}\sum_{i,\alpha\neq\beta} \sum_\sigma \hat{n}_{i\alpha\sigma}\hat{n}_{i\beta,\sigma} + H_{pair} + H_{\mu_d},
\label{hubbard}
\eeq
where $c^\dagger_{i\alpha\sigma}$ creates an electron in a V $d$ orbital $\alpha$ with spin $\sigma$ at site $i$. $J$ is the Hund's coupling, $U'$ the inter-orbital coupling, $\bar{\sigma} = -\sigma$, $\hat{n}_{i\alpha\sigma} = c^\dagger_{i\alpha\sigma} c_{i\alpha\sigma}$, 
and $H_{pair}$ is the pair hopping term defined as
\be
H_{pair}= -\frac{J}{2}\sum_{i,\alpha\neq\beta} \left[c^\dagger_{i\alpha\uparrow} c_{i\alpha\downarrow} c^\dagger_{i\beta\downarrow}c_{i\beta\uparrow} + 
c^\dagger_{i\alpha\uparrow}c^\dagger_{i\alpha\downarrow}c_{i\beta\uparrow} c_{i\beta\downarrow}+ h.c\right].
\ee
Lastly, 
\be
H_{\mu_d} = - \mu^U_d \sum_{i\alpha\sigma} \hat{n}_{i\alpha\sigma},
\ee
is the on-site energy introduced to make $H_U = 0$ at the half-filling condition.\cite{ssreview} It is easy to show that for a five orbital model, $\mu^U_d = (9U - 20 J)/4$ if we adopt the 
form of the Hubbard interactions defined in Eq. \ref{hubbard}.
We will use the values of $U=6$ eV, $J=1$ eV, and $U'=U-2J$ for VO$_2$ and NbO$_2$ throughout this paper.

The slave spin method \cite{ssreview,demedici2005,demedici2009,hassan2010,demedici2011,yu2011,yu2012,yu2013,demedici2014,giovannetti2015,fanfarillo2015,mukherjee2016,pizarro2017,yu2017,leewc2017} will be employed to treat $H=H_{TB} + H_U$.  It is worth mentioning that the slave spin formalism\cite{demedici2005,hassan2010} has been shown to capture the featureless Mott transition in good agreement with DMFT results, and it correctly reproduces the Gutzwiller limit for the quasiparticle weight $Z\to 2x/(1+x)$, where $x$ is the doping away from the half-filling in the large $U$ limit.\cite{hassan2010,yu2012,mukherjee2016} We adopt the $U(1)$ version of the slave spin method described in Ref. [\onlinecite{yu2012}], and focus only on the rutile phase of VO$_2$ in the normal state without any spontaneous symmetry-breaking order in this study. A brief review on the formalism of $U(1)$ slave spin method can be found in Appendix A.

\begin{figure}
\includegraphics{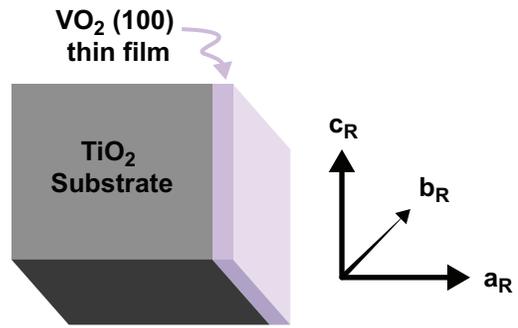}
\caption{\label{fig:100} Schematic illustration of the VO$_2$ thin film grown along the rutile $a$-axis (100) direction. }
\end{figure}
 
The strained case of VO$_2$ thin films grown on a TiO$_2$ substrate with the growth direction along the rutile $a$-axis, denoted as VO$_2$ (100), is schematically illustrated in Fig. \ref{fig:100}. In this case, the lattice constants of the $b$- and $c$-axis of the VO$_2$ match those of the TiO$_2$ substrate while the $a$-axis lattice constant shrinks. Table \ref{table} summarizes the lattice constants used in the present study. We now compare the unstrained (bulk) VO$_2$ and VO$_2$ (100) to resolve the interplay between band structures and electron correlations.

\begin{table}[]
	\caption{\label{table} Lattice constants of the bulk VO$_2$ and VO$_2$ (100) in rutile phase\cite{mukherjee2016}} 
	\begin{ruledtabular} 
		\begin{tabular}{c c c c} 
		System & a$_R$ & b$_R$ & c$_R$ \\ 
		\hline \hline 
		Bulk VO$_2$ & 4.554\AA & 4.554\AA & 2.851\AA \\
                VO$_2$ (100) & 4.47\AA&4.594\AA&2.958\AA\\
		\end{tabular} 
	\end{ruledtabular} 
\end{table}

\section{Results}
\subsection{Non-interacting limit}
\begin{figure}
\includegraphics[width=3.4in]{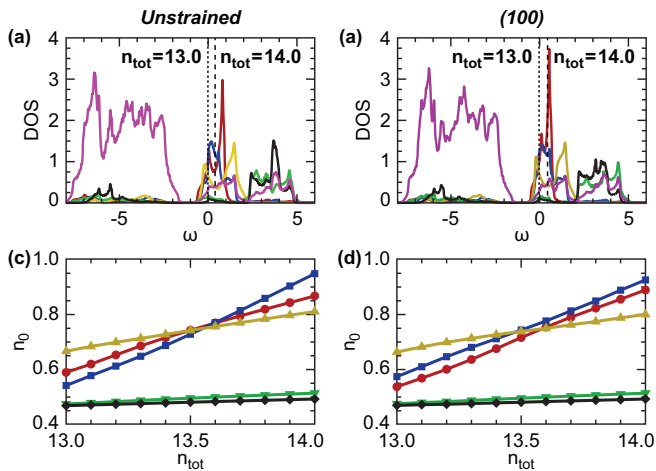}
\caption{\label{fig:non-int} Results of $U=J=0$ for both unstrained (bulk) VO$_2$ and VO$_2$ (100). The density of states on V $d$ and O $p$ orbitals obtained from $H_{TB}$ are plotted for (a) unstrained VO$_2$ and (b) VO$_2$ (100), and the corresponding electron fillings in $d$ orbitals are plotted in (c) and (d) respectively. The figure legends are $d_{x^2-y^2}$ (red, circle), $d_{xz}$ (blue,square), $d_{yz}$ (yellow, up-pointing triangle), $d_{3z^2-r^2}$ (green, down-pointing triangle), $d_{xy}$ (black, diamond). The pink lines in (a) and (b) represent the DOS of the oxygen $p$ orbitals. }
\end{figure}

We first present our results without the interaction term $H_U$. This will serve as a reference to highlight the effects of strain on the band structures. We follow the local coordinate system introduced by Eyert \cite{eyert2002}, in which the $t_{2g}$ orbitals are $d_{x^2-y^2}$ ($d_\parallel$), $d_{xz}$ and $d_{yz}$ ($d_\pi$ bands), and the  $e_g$ orbitals are $d_{3z^2-r^2}$ and $d_{xy}$. Fig. \ref{fig:non-int}(a) summarizes our results for the density of states (DOS) from $H_{TB}$ in unstrained VO$_2$. This result reproduces the previous DFT calculations accurately.\cite{eyert2002} The electron density in the vanadium $d$ and oxygen $p$ orbitals has been found to deviate significantly from the ionic picture, one that predicts full occupation of the oxygen $p$ orbitals, and 1 electron in the vanadium $d$ manifold. While DFT calculations obtain a total electron filling in V $d$ and O $p$ orbitals per unit cell ($n_{tot}$) to be thirteen as expected, the electron filling in the vanadium $d$ orbitals ($n_d$) is significantly larger than 1 due to strong hybridization between V $d$ and O $p$ orbitals. Previous calculations have obtained $n_d$ ranging from $1.0$ to $3.15$,\cite{eyert2002,biermann2005,haverkort2005,weber2012}. Because in DFT-SS we employ a tight-binding model, and the total electron filling can be varied freely by changing the chemical potential $\mu$, we can access general trends in the range of $13 < n_{tot} < 14$, which covers the range of $n_d$ of the experimental interest.

Figs. \ref{fig:non-int}(c) and (d) plot the electron filling in each $d$ orbital as a function of $n_{tot}$ for the unstrained VO$_2$ and VO$_2$ (100). Clearly, in both cases, the electron filling in the $e_g$ orbitals remains almost the same as $n_{tot}$ increases, while those in $t_{2g}$ orbitals increases significantly as $n_{tot}$ increases. This indicates that only the $t_{2g}$ orbitals are important near the Fermi energy, and consequently, the correlation effects will be primarily on $t_{2g}$ orbitals.

Now we discuss the effects of the (100) strain on the band structure. Comparing Fig. \ref{fig:non-int}(a) and (b), it can be seen that under this strain the bandwidth of the $d_{x^2-y^2}$ ($d_\parallel$) near the Fermi energy is significantly reduced, while those of the $d_{xz}$ and $d_{yz}$ ($d_\pi$ bands) remain roughly the same. This orbital-dependent reduction of the bandwidth is one of the crucial ingredients for the occurrence of the orbital selective Mott state driven by strain. 

\subsection{Effects of $H_U$ in unstrained VO$_2$}
Before we turn on the Hubbard interactions $H_U$, it is helpful to discuss the general effects of $H_U$. First, it costs energy to put two electrons on the same site, so electron hopping is suppressed. This results in the reduction of the quasiparticle weight $Z$. Second, $H_U$ could produce orbital-dependent effective on-site potentials that lead to the redistribution of orbital occupation numbers. To see why this happens, we analyze $H_U$ given in Eq. \ref{hubbard}. If $J$ is nonzero, $U' = U-2J < U$. This means that if two electrons are at the same site, the interaction energy will be lower for them to stay in different orbitals than in the same orbital.  As a result, $H_U$ favors an equal orbital filling and this tendency is reflected in the orbital-dependent effective on-site potentials generated by $H_U$. Such an orbital-dependent on-site potential can be ignored only if the Hund's coupling $J$ is negligible, which does not seem to be the case for VO$_2$.\cite{biermann2005,lazarovits2010,weber2012,brito2016}

\begin{figure}
\includegraphics[width=3.4in]{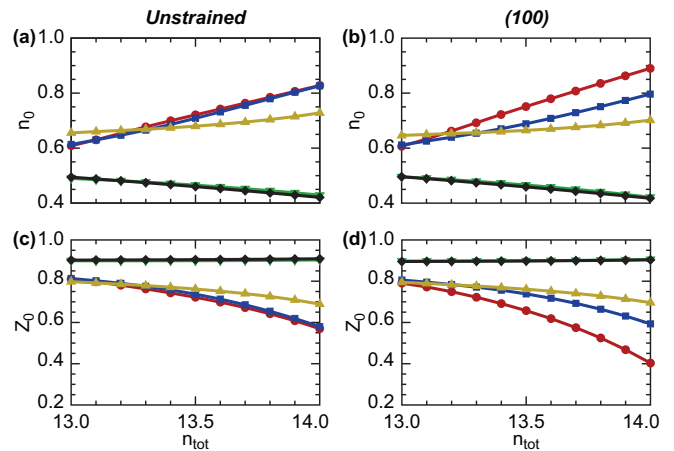}
\caption{\label{fig:u6jh1} The electron fillings of $d$ orbitals with $U=6$ eV and $J = 1$ eV are calculated using the slave-spin method for (a) unstrained VO$_2$ and (b) VO$_2$ (100), and 
the corresponding quasiparticle weights are plotted in (c) and (d) respectively. The figure legends are $d_{x^2-y^2}$ (red, circle), $d_{xz}$ (blue,square), $d_{yz}$ (yellow, up-pointing triangle), $d_{3z^2-r^2}$ (green, down-pointing triangle), $d_{xy}$ (black, diamond).}
\end{figure}

The slave-spin method can capture both effects from $H_U$, as demonstrated in previous works.\cite{yu2012,yu2013,mukherjee2016,yu2017,leewc2017} The orbital occupation numbers and the quasiparticle weights as a function of the total electron filling $n_{tot}$ with $U=6$ eV and $J=1$ eV for the unstrained VO$_2$ are plotted in  Figs. \ref{fig:u6jh1}(a) and (c). It can be seen that the Hubbard interactions make the electron occupation of the $t_{2g}$ bands more symmetrical. Moreover, the quasiparticle weights of $t_{2g}$ bands are much smaller than those of the $e_g$ bands. These observations are consistent with the discussions given above. It is also interesting to note that the correlation effects are more and more pronounced as $n_{tot}$ approaches 14.  As mentioned above, since the $e_g$ bands are away from the Fermi energy, they are not affected by $H_U$, and most extra electrons will go into the $t_{2g}$ bands. Consequently, the electron filling in the $t_{2g}$ bands increases with $n_{tot}$. If the electron filling in the $t_{2g}$ bands ($n_{t_{2g}}$) approaches 3, the `half filling' condition for a three-orbital ($t_{2g}$) model, the correlation effects in the $t_{2g}$ bands will be enhanced significantly. From Fig. \ref{fig:u6jh1}(a), it can be seen that $n_{t_{2g}}$ increases from 1.88 to 2.38 as $n_{tot}$ increases from 13 to 14, which explains the rapid reduction of the quasiparticle weights in the $t_{2g}$ bands.
 
\subsection{Orbital Selective Mott State in VO$_2$ (100)}
Here, we discuss the strained case of VO$_2$ (100) with Hubbard interactions. As mentioned above, the strain results in an orbital-dependent reduction of the bandwidth, and turning on the $H_U$ interaction leads to even more non-trivial effects, as shown in Figs. \ref{fig:u6jh1}(b) and (d). An intriguing difference in the behaviors of the quasiparticle weights can be seen from the comparison between strained and unstrained cases. As $n_{tot}$ is fixed, the quasiparticle weights of the $d_\pi$ bands remain roughly the same but the $d_\parallel$ band decreases significantly in the strained system.  This suggests only the $d_\parallel$ band is driven to be more correlated under strain. Moreover, we find that the electron filling in $d_\parallel$ band increases, while that in $d_\pi$ bands decreases in VO$_2$ (100). These two behaviors strongly indicate that strained VO$_2$ is approaching an orbital selective Mott transition (OSMT), consistent with our previous reports \cite{mukherjee2016,quackenbush2016} 
 
\begin{figure}
\includegraphics[width=3.4in]{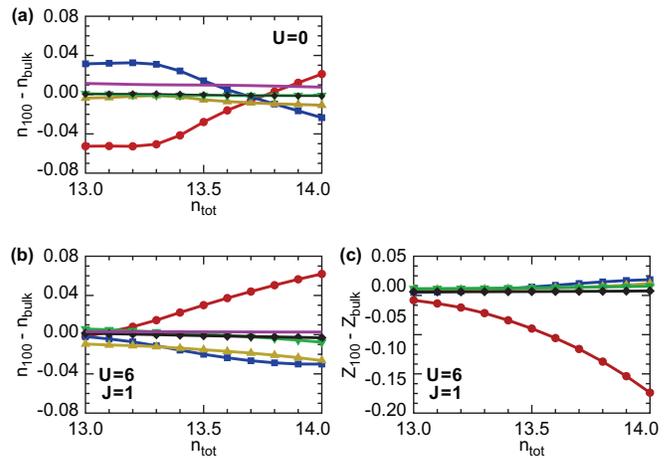}
\caption{\label{fig:delta-nz} (a) Difference in the electron filling in the non-interacting limit ($U=0$) on each $d$ orbital between VO$_2$ (100) and the unstrained VO$_2$. 
Differences in the electron filling (b) and the quasiparticle weight (c) on each $d$ orbital between VO$_2$ (100) and the unstrained VO$_2$ for $U=6$ eV and $J=1$ eV.
The figure legends are $d_{x^2-y^2}$ (red, circle), $d_{xz}$ (blue,square), $d_{yz}$ (yellow, up-pointing triangle), $d_{3z^2-r^2}$ (green, down-pointing triangle), $d_{xy}$ (black, diamond).}
\end{figure}

Fig. \ref{fig:delta-nz} plots the differences in electron filling and quasiparticle weights between VO$_2$ (100) and unstrained VO$_2$. In the non-interacting limit, shown in Fig. \ref{fig:delta-nz}(a), we find that $\Delta n$ in the $d_\parallel$ (red dots) changes sign as $n_{tot}$ increases from 13 to 14. With the interactions, it can be seen in Figs. \ref{fig:delta-nz}(b) and (c), that $\Delta n$ ($\Delta Z$) in the $d_\parallel$ increases (decreases) monotonically as $n_{tot}$ increases from 13 to 14. This observation proves that electron correlation plays a crucial role here, and the mechanism for the OSMT relies on the cooperative interplay between strain-modulated band structures and electron correlation. In VO$_2$ (100), because the $c$-axis is elongated due to the strain, and the $d_\parallel$ band has a large wavefunction overlap along $c$-axis, the bandwidth of the $d_\parallel$ band is significantly reduced compared to the $d_\pi$ bands. Since correlation effects are usually characterized by the ratio of the Hubbard interaction to the bandwidth ($U/W$), the $d_\parallel$ band effectively becomes more correlated if Hubbard interactions exist. These two strain effects cooperatively induce the intriguing OSMT in VO$_2$ (100).

\subsection{Single particle spectral function}

\begin{figure*}
\includegraphics{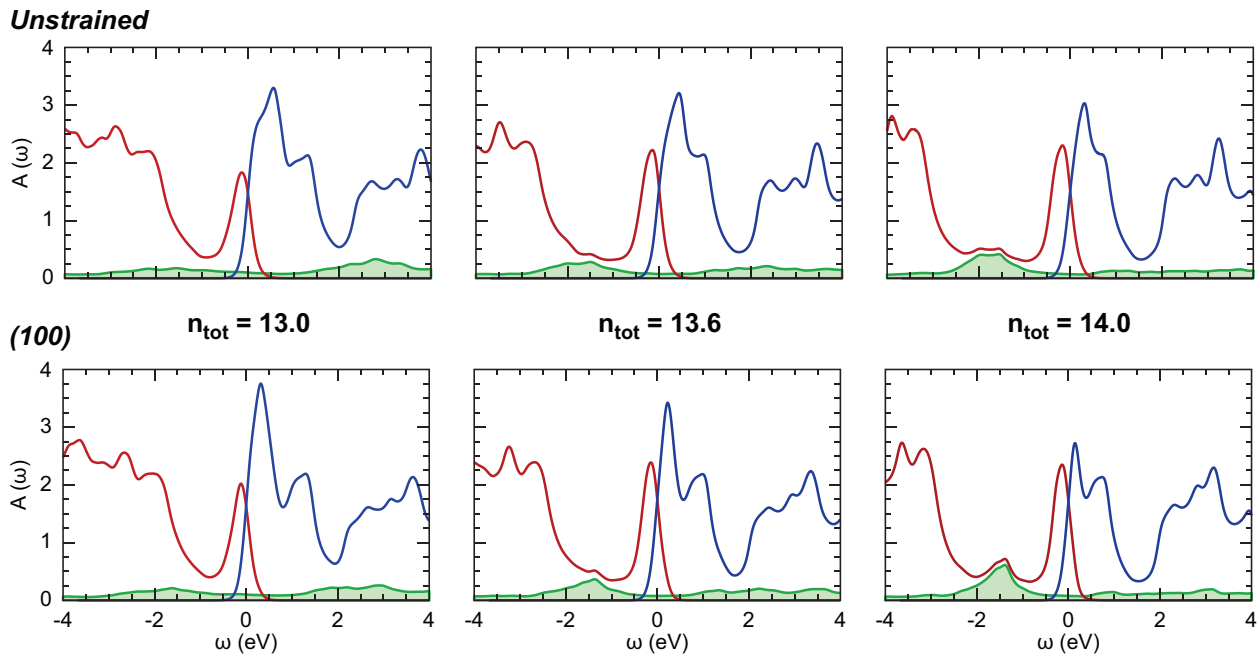}
\caption{\label{fig:aw} Simulations of spectral functions for occupied states (red) and unoccupied states (blue) using DFT-SS method. The green line represents the incoherent
spectrum due to the lower and upper Hubbard band.}
\end{figure*}

Another important feature of Mott physics is the emergence of the incoherent spectra, known as the upper and lower Hubbard bands (LHB and UHB), in the single particle spectral function.\cite{phillips2010,leewc2011} 
We adopt the formalism of the single particle Green's function in the slave-spin method derived in Ref. [\onlinecite{yu2011}], 
and the spectral function at the mean-field level can be calculated directly by
\be
A(\omega) = \sum_k \sum_{\alpha \sigma} A_{\alpha,\sigma}(\vec{k},\omega),
\ee
where $A_{\alpha,\sigma}(\vec{k},\omega)$ is the spectral function with orbital $\alpha$ and physical spin $\sigma$\cite{yu2011}.
In the slave spin formalism, the electron Green's function can be approximated as the convolution of the spinon and slave spin propagators, and 
contributions from the coherent and incoherent parts can be separately calculated.\cite{yu2011}
The derivation of $A_{\alpha,\sigma}(\vec{k},\omega)$ in the slave spin method can be found in Appendix B. 
With the spectral function, we can simulate spectral functions for occupied states and unoccupied states by
\beq
A^{occu}(\omega) &=& n_F(\omega) A(\omega),\nn\\
A^{unoccu}(\omega) &=& [1-n_F(\omega)] A(\omega).
\eeq
$n_F(\omega)$ is the Fermi Dirac function at room temperature $\frac{1}{\beta} =25$ meV. The photoemission spectra is typically proportional to $A^{occu}(\omega)$, the spectral weights of occupied states. Fig. \ref{fig:aw} presents the simulated spectra for cases of unstrained and VO$_2$ (100) with several different total electron filling numbers. At smaller $n_{tot}$, the incoherent spectrum (green line) due to the LHB around 1 eV in energy can be clearly seen, and becomes more pronounced in the strained sample. This is more evidence of strong correlation effects in the strained sample, consistent with the OSMT scenario. At larger $n_{tot}$, the incoherent spectrum is pushed to more negative energy, and both bulk and strained samples have a pronounced LHB. This is not surprising, since as $n_{tot}$ approaches 14, the electron filling in the $t_{2g}$ orbitals approach 3, which is `half-filling' for a three-orbital model. In this case, the system is already close to the Mott limit, even for the unstrained material. As a result, the physical properties do not change dramatically under the strain.

\section{\bf Comparison with Resonant Photoemission Data of Strained VO$_2$}
\begin{figure}
\includegraphics[width=3.0in]{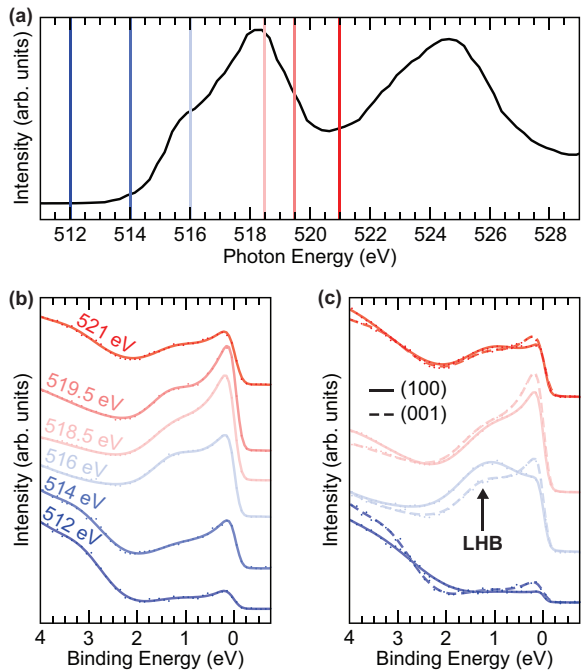}
\caption{\label{fig:VO2_RPES} (a) Vanadium L-edge XAS of the high temperature rutile phase of VO$_2$/TiO$_2$ (001). The photon energies used to excite resonant photoemission are denoted by vertical lines. (b) The corresponding RPES spectra for VO$_2$/TiO$_2$ (001) in the rutile phase. (c) Direct comparison of the RPES on (001) and (100) rutile phases for select photon energies, showing the strain-enhancement of the incoherent features at $\sim$ 1.5 eV. The arrow highlights the enhancement of the LHB with strain.}
\end{figure}

\begin{figure}
\includegraphics[width=3.5in]{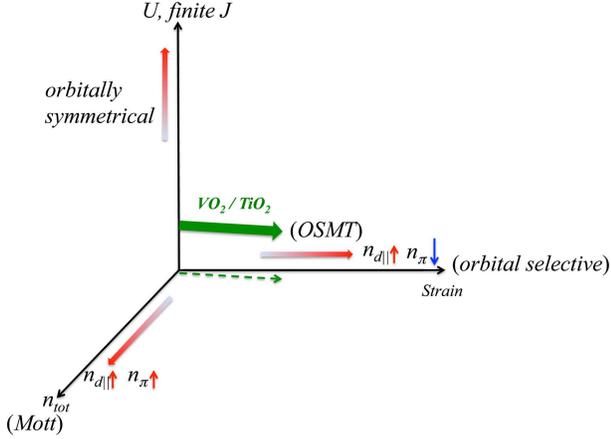}
\caption{\label{fig:phase} Schematic illustration of the orbital-dependent redistributions of electron fillings due to the strain and the electron correlation. In the epitaxial
VO$_2$/TiO$_2$ (100)\cite{mukherjee2016}, the system is pushed toward the orbital selective Mott transition (OSMT), as indicated by the
green arrow.}
\end{figure}

We previously presented evidence of a strain-induced OSMT for epitaxial VO$_2$/TiO$_2$ films in the high temperature, rutile phase using a combination of HAXPES and V L$_3$-edge XAS.\cite{mukherjee2016} The preferential orbital filling was evidenced by the XAS dichroism and the loss of quasiparticle weight was observed in the HAXPES.  In addition, $c$-axis elongation also increased spectral weight away from the Fermi energy (i.e., at $\sim$ 1.5 eV).  Our simulations presented in Fig. \ref{fig:aw} suggest that the LHB contributes to this region and is dependent on the strain orientation. In order to interrogate these features further, we have employed {\it{resonant}} x-ray photoemission spectroscopy (RPES) at the V L$_3$-edge to enhance our sensitivity of the occupied V 3$d$ orbital region. The 10 nm epitaxial films were grown by reactive molecular beam epitaxy at Cornell University, further details are provided by Paik et al.\cite{paik2015}  Fresh samples were prepared ahead of the RPES experiments at the 29-ID beamline at Argonne Photon Source on the soft x-ray ARPES end-station. The samples were measured in their respective high temperature phases (at room temperature for (001) and at 398K for (100)), and spectra were energy calibrated using a combination of Au foil and VO$_2$ film references. When at resonance, the direct recombination of the decay process emits 3$d$ electrons when excited at the L-edge resulting in enhanced 3$d$ contributions in the measured x-ray photoemission spectra. Eguchi {\it{et al.}} previously employed V L-edge RPES to measure the insulating and metallic phases of 10 nm thin VO$_2$ (001) films, where they were able to enhance sensitivity to the quasiparticle peak and incoherent feature.\cite{eguchi2008}  For the VO$_2$ (001) case we reproduced the metallic phase RPES spectra presented in Ref. \cite{eguchi2008} for the same energies they employed, as shown in Fig. \ref{fig:VO2_RPES}(b). Fig.  \ref{fig:VO2_RPES}(c) directly compares the resonant spectra for the VO$_2$ (001) and (100) cases for select energies. The largest enhancement compared to the off-resonance data was observed for the energies resonant with the knee and main peak of the V L$_3$-edge absorption line. We note that these energies displayed the largest strain-induced dichroism due to the preferential orbital filling associated with the strain-induced OSMT.\cite{mukherjee2016} In addition to the weaker quasiparticle peak intensity of the VO$_2$ (100) compared to the (001) case that we reported on previously,\cite{mukherjee2016} the increased sensitivity afforded by the resonance reveals how this effect is correlated with increased incoherent peak at $\sim$ 1.5 eV for the (100) case.  These data agree with simulated spectral functions in Fig. \ref{fig:aw}, where a stronger incoherent peak at $\sim$ 1.5 eV is always observed for the case where the c-axis is elongated i.e. (100), compared to bulk or when the c-axis is compressed i.e (001). As a result, we can assign this feature as the lower Hubbard band and conclude that VO$_2$ can be strain-engineered as predicted by our modeling of the OSMT.  Fig. \ref{fig:phase} summarizes the general trends of the redistribution of electron occupation numbers due to the interplay between strain and Hubbard interactions in the epitaxial VO$_2$/TiO$_2$ systems.

\section{Strain effect on N\MakeLowercase{b}O$_2$ in rutile phase}
\begin{figure}
\includegraphics[width=3.4in]{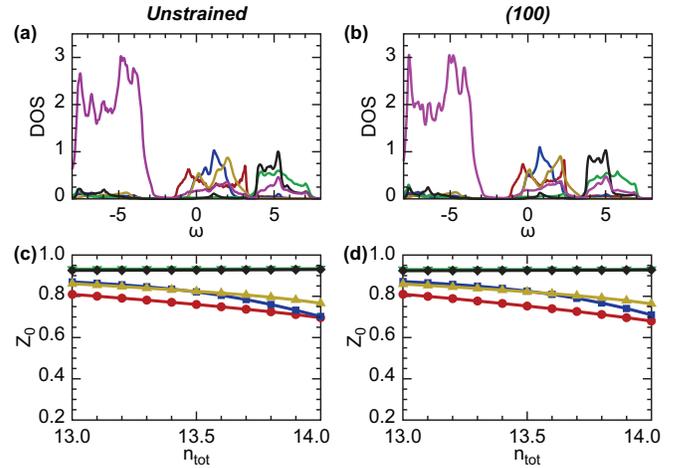}
\caption{\label{fig:nbo2} The non-interacting DOS of $d$ and $p$ orbitals for (a) unstrained NbO$_2$ and (b) NbO$_2$ (100) in rutile phase. The orbital-dependent quasiparticle weights
with $U=6$ eV and $J = 1$ are plotted in (c) and (d) respectively.
The figure legends are $d_{x^2-y^2}$ (red, circle), $d_{xz}$ (blue,square), $d_{yz}$ (yellow, up-pointing triangle), $d_{3z^2-r^2}$ (green, down-pointing triangle), $d_{xy}$ (black, diamond).}
\end{figure}

As a comparison, in this section we study the strain effect on NbO$_2$ in rutile phase. NbO$_2$ undergoes a metal-to-insulator transition at a temperature around 810$^\circ C$. It simultaneously undergoes a crystal structure change between a high temperature, undistorted rutile phase and a low temperature, body-centered tetragonal, distorted rutile phase.\cite{ohara2015} Although strain engineering on NbO$_2$ in the rutile phase has not yet been achieved experimentally, here we investigate the change of quasiparticle weights in rutile NbO$_2$ under the same strain condition as the VO$_2$ case. For the bulk NbO$_2$, we adopt the LDA-optimized lattice parameters, $(a_R,b_R,c_R)=(4.93\AA,4.93\AA,2.9\AA)$\cite{ohara2015}. For NbO$_2$ (100), we choose the lattice parameters of $(a'_R,b'_R,c'_R)=(4.74\AA,4.93\AA,3.016\AA)$, which is in the same strain condition of VO$_2$ (100) studied in previous sections. Note that the spin-orbit coupling is not included in our calculations. Fig. \ref{fig:nbo2}(a) and (b) plot the non-interacting DOS obtained from the DFT-fitted $H_{TB}$ for bulk and (100) respectively, and the orbital-dependent quasiparticle weights calculated from our DFT-SS method with $U=6$ eV and $J = 1$ are summarized in Fig. \ref{fig:nbo2}(c) and (d). We find that although the bandwith of the $d_{x^2-y^2}$ orbital still has a significant reduction in (100) compared to the bulk, the quasiparticle weights are almost the same in NbO$_2$ with and without the strain, which is very different from the VO$_2$ case. We attribute this result to the fact that the bandwidth of $d$ orbitals in NbO$_2$ is larger than that of VO$_2$, thus, the ratio of the interaction energy to the bandwidth ($U/W$) is generally smaller even with the same values of interaction parameters. This suggests that NbO$_2$ has a weaker correlation and consequently, is not an ideal material for modulating Mott physics with strain.

\begin{figure}
\includegraphics{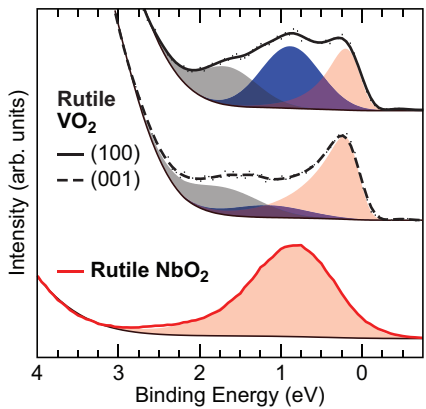}
\caption{\label{fig:vo2-nbo2} HAXPES valence band edge spectra of rutile VO$_2$ (100), VO$_2$ (001), and relaxed NbO$_2$ (110) thin films with peak fits. All the measurements were performed above their respective MIT temperatures. Compared to the VO$_2$ thin films, the spectrum of 
NbO$_2$ can be well-fit by a single peak, indicating the absence of the lower Hubbard band (LHB).}
\end{figure}

To experimentally investigate this conclusion, HAXPES was performed on rutile VO$_2$ (100), VO$_2$ (001), and relaxed NbO$_2$ (110) thin films, as shown in Fig. \ref{fig:vo2-nbo2}. 
The HAXPES measurements were performed with 5.95 keV x-rays at the I09 beamline of the Diamond Light Source. HAXPES spectra were energy-resolved and measured using a Scienta Omicron EW4000 high-energy analyzer with a $\pm 30^\circ$ acceptance angle. The x-ray beam was monochromatized using a Si (111) double-crystal monochromator followed by a Si(004) channel-cut crystal
resulting in an overall energy resolution of  $\leq 250$ meV. The films were all heated above their respective MIT temperatures during measurement to transition them into their metallic, rutile phases. It can be clearly seen that unlike the VO$_2$ thin films, the Nb $d$-orbital feature near the Fermi level of NbO$_2$ can be well-fit by a single peak without the need for an additional hump feature associated with the lower Hubbard band. This observation is completely consistent with our conclusion on the NbO$_2$ discussed above.

\section{Conclusion}
In this paper, we have studied general trends in the redistribution of electron occupation numbers due to interplay between strain and correlations in epitaxial VO$_2$/TiO$_2$ systems, which is summarized in Fig. \ref{fig:phase}. Since the $e_g$ bands are far from the Fermi energy, the correlation effects will be related mainly to the $t_{2g}$ bands. We find that stronger Hubbard interaction tends to make the electron fillings in $t_{2g}$ orbitals more even, which is a general trend for a non-zero Hund's coupling $J$. Moreover, because most extra electrons will go into the $t_{2g}$ orbitals as $n_{tot}$ increases, increasing $n_{tot}$ will push the system closer to the half-filling condition ($n=3$ for a three-orbital $t_{2g}$ system). This consequently enhances correlation effects, resulting in the reduction of quasiparticle weights, and emergence of the incoherent spectrum as shown in our calculations. Finally, the strain in VO$_2$ (100) causes an elongation of the rutile $c$-axis that produces a significant reduction of the bandwidth, particularly on the $d_{\parallel}$ band, but not on the $d_\pi$ bands. This orbital-dependent reduction of the bandwidth leads to an intriguing `orbital-selective' Mott transition (OSMT) in which only one of the orbitals is pushed much closer to the Mott insulating state. Based on our results, we conclude that in epitaxial VO$_2$/TiO$_2$ (100) and (110),\cite{mukherjee2016} the system moves toward the OSMT due to $c$-axis elongation by the strain from lattice matching with the TiO$_2$ substrate, as indicated by the green arrow in Fig. \ref{fig:phase}. 

In summary, we have employed density functional theory combined with the slave spin method (DFT+SS) to study the electronic structures of epitaxial VO$_2$ films under strain in the rutile phase. We have found that while asymmetrical orbital occupation numbers are favored due to band structure effects in the strained systems with an elongated $c$-axis, strong electron-electron correlation drives orbital-dependent modifications of the quasiparticle weights, as well as the incoherent spectra in the single particle Green's function. The interplay between these two distinct effects pushes epitaxial VO$_2$ films toward an orbital selective Mott transition (OSMT) in the rutile phase without a Peierls instability. Our results indicate that Mott physics is important and can be significantly modulated even in the rutile phase of VO$_2$ by appropriate strain-engineering.

\section{Acknowledgement}
This work is supported by the Air Force Office of Scientific Research under award number FA9550-18-10024.
This research used resources  of the Advanced Photon Source, a U.S. Department of Energy (DOE) Office of Science User Facility operated by the DOE Office of Science by Argonne National Laboratory under Contract No. DE-AC02-06CH11357; additional support by National Science Foundation under Grant no. DMR-0703406. Part of calculations used the Extreme Science and Engineering Discovery Environment (XSEDE) supported by National Science Foundation grant number ACI-1548562.  AR acknowledges the support from Imperial College London for her Imperial College Research Fellowship.  We thank Diamond Light Source for access to beamline I09 (SI20647 and SI21430-1) that contributed to the results presented here. J.E.B acknowledges funding by the Leibniz association within the Leibniz Competition through a project entitled 'Physics and control of defects in oxide films for adaptive electronics'.

\section{Appendix A: U(1) slave spin formalism}
The motivation behind the slave-spin formalism is to introduce a two-level degree of freedom to handle the charge dynamics of interacting fermions. The occupation of each local degree of freedom, whether it be site, spin, or orbital, can be treated as either occupied (1), or unoccupied (0), so a spin algebra can be used to represent this. That means, the spin operator seen in the following derivations is not for the physical spin, but rather, it is a `slave' spin capturing the dynamics of `charge' degrees of freedom. The physical spin is described by fermionic operators called spinons. With these definitions, one can rewrite the electron creation operator as:
\be
c^\dagger_{i\alpha\sigma}\equiv S^+_{i\alpha\sigma} f^\dagger_{i\alpha\sigma},
\ee
where $i$ is the site index, $\alpha$ is the orbital index, $\sigma$ is the physical spin. The spinon $f^\dagger_{i\alpha\sigma}$ is a fermionic operator capturing the dynamics of the physical spin. Because the Hilbert space for the charge degree of freedom is parameterized as two levels (states with 0 and $-e$ charge), which is fully described by a spin 1/2 algebra, we can have $S^+_{i\alpha\sigma}$  be the spin raising operator for the charge part of the electrons. Since we have a slave spin for each set of quantum numbers $(\alpha,\sigma)$, $S^+_{i\alpha\sigma}$ carries both the orbital and the physical spin indices. Similar to other slave particle formalisms, the Hilbert space has been enlarged in this process, and will contain unphysical states. For example, a state such as $| n^{f}_{i \alpha \sigma} = 0, S^{z}_{i \alpha \sigma} = +1/2 \rangle$ would be unphysical, and must be projected out of the enlarged space. This projection can be done by enforcing the constraint 
\be
S^z_{i\alpha\sigma} = f^\dagger_{i\alpha\sigma}f_{i\alpha\sigma} - \frac{1}{2}.
\ee
Following the procedure outlined in Ref. [\onlinecite{yu2012}], we can define the mean-field parameters 
\be
z_{\alpha,\sigma} \equiv \langle S^+_{i\alpha\sigma}\rangle,
\ee
and $\lambda_{\alpha,\sigma}$ being the Lagrange multipliers to enforce the constraint on the average over the sites:
\be
\frac{1}{N}\sum_i \langle S^z_{i\alpha\sigma}\rangle = \frac{1}{N} \sum_i \left(\langle f^\dagger_{i\alpha\sigma}f_{i\alpha\sigma}\rangle - \frac{1}{2}\right),
\ee
where $N$ is the total number of sites.
At the mean-field level, we have the decoupled Hamiltonians for the spinons ($H^f$) and for the slave spins ($H^S$) which can be used to 
solve $\{z_{\alpha,\sigma},\lambda_{\alpha,\sigma}\}$ self-consistently. After the self-consistent solution is obtained, the quasiparticle weight for each orbital can be evaluated by
\be
Z_{\alpha,\sigma} = \vert z_{\alpha,\sigma}\vert^2.
\ee
Because we are only interested in paramagnetic phase, we have $Z_{\alpha} = \vert z_{\alpha,\uparrow}\vert^2 = \vert z_{\alpha,\downarrow}\vert^2$.

\section{Appendix B: Single Particle Green function}
Using the same procedure given in Ref. [\onlinecite{yu2011}], the imginary-time ordered Green function can be evaluated at the mean-field level as
\beq
G_{\alpha,\sigma}(\vec{k},\tau) &\equiv& -\langle T_\tau c_{\vec{k}\alpha\sigma}(\tau) c^\dagger_{\vec{k}\alpha\sigma}(0)\rangle\nn\\
&\approx& -\langle T_\tau S^-_{\alpha\sigma}(\tau) S^+_{\alpha\sigma}(0)\rangle\langle T_\tau f_{\vec{k}\alpha\sigma}(\tau) f^\dagger_{\vec{k}\alpha\sigma}(0)\rangle\nn\\
\label{green0}
\eeq

In the Lehmann representation, the retarded Green function can be evaluated directly by
\beq
G^{ret}_{\alpha,\sigma,\lambda}(\vec{k},\omega) &=& \frac{1}{N}\sum_{n,m}
\vert\langle n\vert S^+_{\alpha\sigma}\vert m\rangle\vert^2 \big(U_{\vec{k}\sigma}^{\alpha\lambda}\big)^*U_{\vec{k}\sigma}^{\alpha\lambda} \nn\\
&\times& \frac{\big[e^{-\beta E_m}(1-n^f_\lambda(\vec{k})) + e^{-\beta E_n} n^f_\lambda(\vec{k})\big]}{\omega+i\eta - (E_n - E_m) - \epsilon_{\lambda}(\vec{k})},
\label{green}
\eeq
where $\{\vert m\rangle\}$ and $\{ E_m\}$ are the sets of eigenvectors and eigenvalues of the slave spin
mean-field Hamiltonian $H^S$, $\lambda$ is the band index, $n^f_\lambda(\vec{k})$ is the Fermi distribution function for the state with momentum $\vec{k}$ in $\lambda$th band, 
$\{\epsilon_{\lambda}(\vec{k})\}$ is the set of eigenvalues of the spinon mean-field Hamintonian
$H^f$, and $U_{\vec{k}\sigma}^{ij}$ is the unitary matrix that diagonalizes $H^f$. 
Eq. \ref{green} can be evaluated directly without any difficulty after the mean-field equations are solved.

With the electron Green function given in Eq. \ref{green}, the electron spectral function for electrons in each orbital can be obtained straightforwardly by 
\be
A_{\alpha,\sigma}(\vec{k},\omega) = - 2 \sum_{\lambda}\Im G^{ret}_{\alpha,\sigma,\lambda}(\vec{k},\omega).
\ee

One common problem with the slave particle approaches is the missing spectral weight at doping away from the half-filling.
We check the sum rule:
\be
\int_{-\infty}^\infty d\omega A_{\alpha,\sigma}(\vec{k},\omega) = 2n_{\alpha\sigma} - 2(2n_{\alpha\sigma} -1)\left[1-n^f_{\alpha\sigma}(\vec{k})\right],
\label{sumgreen}
\ee
and we find that the integrated spectral weight equals to 1 only as $2n_{\alpha\sigma} -1=0$, which corresponds to the 'half-filling' in the orbital $\alpha$.
This failure is an artifect of the mean-field approximation in which the couplings between spinons and slave spins are neglected in Eq. \ref{green0}.
It is generally expected that once the couplings are turned on, the spinon Fermi surface will be smeared out. 
As a result, we can set $n^f_{\sigma\vec{k}} = 1/2$ in Eqs. \ref{green} and \ref{sumgreen} at the mean-field level, which leads to the correct sum rule of
\be
\int_{-\infty}^\infty d\omega A_{\alpha,\sigma}(\vec{k},\omega) = 1.
\ee 

Finally, as pointed out in Ref. [\onlinecite{yu2011}], the contribution to the spectral function from $m=n$ in Eq. \ref{green} is the coherent part, while 
other contributions from $m\neq n$ corresponds to the incoherent part.

\end{document}